# Phase control of a z-current driven plasma column

S. Kawata[1*], T. Karino[1], Y. J. Gu[2, 3]

[1]Graduate School of Engineering, Utsunomiya University, Utsunomiya, Japan
[2] Institute of Physics of the ASCR, ELI-Beamlines, Na Slovance 2, 18221 Prague, Czech Republic
[3] Institute of Plasma Physics of the CAS, Za Slovankou 1782/3, 18200 Prague, Czech Republic

A dynamic mitigation is presented for sausage and kink instability growths of a z-current driven magnetized plasma column. We have proposed a dynamic mitigation method based on a phase control to smooth plasma non-uniformities and to mitigate the instability growth in perturbed plasma systems. In this paper we found that a wobbling motion of the z-current electron axis induces a phase-controlled perturbation, so that the growths of the sausage and kink instabilities are successfully mitigated. In general, plasma instabilities emerge from perturbations, and the perturbation phase is normally unknown. However, if the perturbation phase is known or actively imposed by, for example, a designed driver wobbling behavior, the instability growth would be controlled and mitigated by a superimposition of the perturbations imposed. The results in this paper demonstrate that the wobbling z-current electron beam would provide an improvement in the plasma column stability and uniformity.

PACS number: 52.40.Mj, 52.59.-f, 52.25.Xz, 47.20.-k,

* Corresponding author Email address: kwt@cc.utsunomiya-u.ac.jp

A dynamic mitigation mechanism was proposed in Refs. 1-5 for plasma instabilities and non-uniformities. In this paper the dynamic smoothing method is applied to a *z*-current driven plasma column, and 3D particle-in-cell simulations are performed to investigate the dynamic mitigation effect on the sausage and kink instabilities of the *z*-current driven plasma column[6-10]. In our study the plasma column is sustained by the *z* electron current, and the *z*-electron beam axis rotates around the column axis. The results in this paper demonstrate that the wobbling z-current electron beam would provide an improvement in the plasma column stability and uniformity.

In order to control physical systems, like tall buildings, the feedback control[11] is employed widely to stabilize the tall buildings. In the feedback control the perturbation amplitude and phase are measured, and another perturbation with the reverse phase is applied actively to compensate the original perturbation. In plasmas we cannot measure the amplitude and the phase of the plasma perturbations. Therefore, the instability growth rate is usually discussed in plasmas. However, if the perturbation phase is actively imposed by a wobbling or oscillation driving source, the amplitude of the perturbation can be controlled in the same way[1-5] as the usual control theory.

In Refs. 1, 2 and 4 we have proposed the dynamic smoothing mechanism based on a phase control and applied it to reduce the Rayleigh-Taylor instability (RTI) growth by using the beam axis wobbling. References 2 and 12 also showed that the dynamic smoothing mechanism would be rather robust against the perturbation in the driver dynamic behavior. The smoothing mechanism was also applied to the filamentation instability in Ref. 3. The driving particle-beam axis wobbling mitigates the growth of the filamentation instability[13-16]. The oscillating beam induces the phase-defined continuous perturbations. The growth of the integrated instability amplitude is mitigated. In heavy ion inertial fusion (HIF) the heavy ion beam (HIB) axis can be also wobbled in the heavy ion accelerator with a high frequency[17-20]. In HIF the amplitude and phase of the perturbation applied are defined by the HIB axis oscillation behavior. The dynamic smoothing mechanism was also applied to smooth the HIBs deposition energy non-uniformity and improve the fuel target implosion uniformity[5, 12]. In Ref. 5 we found that the HIBs wobbling behavior controls the implosion acceleration oscillation and also contributes to reduce the implosion non-uniformity. In Ref. [21], this dynamic stabilization mechanism is applied to the two-stream instability stabilization, in which the classical two-stream instability driven by a constant relative drift velocity is modified by the additional oscillation on the relative velocity. The time-dependent drift velocity opens a new stable window in the two-stream instability.

In this paper we propose to use the control theory to mitigate the sausage and kink instability growths of the *z*-current driven plasma column. The study would contribute to realize more stable magnetized plasmas[6, 7, 27] and also to understand the behavior of jets created in space[28]. First, we shortly summarize the dynamic smoothing mechanism in plasmas. Then the dynamic mitigation results are presented for the sausage and kink instability growths. The results demonstrate that the dynamic smoothing mechanism would improve the plasma column uniformity by the wobbling driver electron beam.

In Refs. [22-24] one dynamic stabilization mechanism was proposed to stabilize the RTI based on a strong oscillation of acceleration with a large oscillation amplitude. In this mechanism, the total acceleration oscillates strongly, and the additional oscillating force is added to create a new stable window in the system. Originally this dynamic stabilization mechanism was proposed by P. L. Kapitza [25], and it was applied to stabilize an inverted pendulum. The inverted pendulum is an unstable system, and a strongly and rapidly oscillating acceleration is applied on the system in Ref. [25], and the inverted pendulum system has a stable window. In this case, the equation for the unstable system is modified, and has another force term coming from the oscillating acceleration. In this mechanism, the growth rate is modified by the strongly oscillating acceleration. When the inverted pendulum is subjected by a strongly oscillating acceleration $A \sin \omega t$, we obtain the following Mathieu-type equation [26] for the oscillating angle of $\theta$(t): $d^2\theta(t)/dt^2 = (g/l)\,\theta(t) - A\omega^2\,\theta(t)\sin\omega t$. Here *l* is the length of the pendulum. When *A*=0, the inverted pendulum becomes unstable. However, the second term of the righthand side is added to the system, and stable windows appear in the inverted pendulum system [25]. The stability condition is described as $A - 0.5 < 2g/(l\omega^2) < A^2$.[25, 26] The stability condition shows that the additional acceleration oscillation at the second term of the righthand side should be fast enough, and the amplitude of $A$ must satisfy the stability condition. The dynamic stabilization mechanism works on the inverted pendulum. It could be applied to stabilize plasmas, though it would be difficult to apply this mechanism to our tall buildings, bridges or large structures in our society.

On the other hand, there is another control mechanism based on the phase control[1-5]. Figure 1 shows the dynamic mitigation mechanism. Let us now consider a perturbed plasma system, which has a single mode of $k = 2\pi/\lambda$ with the amplitude of $a = a_0 e^{ikx+\gamma t}$. For a stable system $\gamma$ is negative, and for an unstable system $\gamma$ means the instability growth rate. Here $\lambda$ is the wave length. Figure 1(a) shows an example initial perturbation in an unstable system. The initial perturbation in the unstable system is assumed to be imposed at *t*=0, and the perturbation grows with the growth rate of $\gamma$. If the next perturbation is actively superimposed on the system at $t=\Delta t$, and also if the perturbation added has the inverse phase as shown in Fig. 1(b), the integrated amplitude growth is mitigated (see Fig. 1(c)). An ideal dynamic smoothing mechanism is demonstrated in Fig. 1[1-5]. It is difficult to detect the perturbation phase and amplitude in plasmas. However, as presented in Fig. 1, if the energy driver, which may have perturbations, provides a wobbling motion, we could expect a control of the perturbation amplitude growth. A superimposed perturbation for a physical quantity $F$ at $t=\tau$ may be expressed as follows: $F = \delta F e^{i\Omega\tau} e^{\gamma(t-\tau)+i\vec{k}\cdot\vec{x}}$. Here we assume the uniform oscillation of the perturbed driver in time. Here the amplitude is described by $\delta F$, $\Omega$ shows the wobbling frequency of the driving beam, and $\Omega\tau$ is the phase shift of the perturbations superimposed. The integrated actual perturbation at $t$ is derived as follows: $\int_0^t d\tau\ \delta F e^{i\Omega\tau} e^{\gamma(t-\tau)+i\vec{k}\cdot\vec{x}} \propto \frac{|\gamma|+i\Omega}{\gamma^2+\Omega^2} \delta F e^{\gamma t} e^{i\vec{k}\cdot\vec{x}}$. When $\gamma \leq 0$, the system is stable and the equation shows a simple dynamic smoothing of the perturbations. When $\gamma \geq 0$ and $\Omega \gg \gamma$, the system is unstable and the amplitude reduction ratio is ${}^{\gamma}/_{\Omega}$ [1, 2]. Even for $\Omega \cong \gamma$ we can still expect the significant mitigation. At this point, it should be noted that the integrated perturbation amplitude is mitigated well, but the growth rate $\gamma$ of the instability does not change. The result suggests that the wobbling frequency $\Omega$ should be high compared with the instability growth rate of $\gamma$ for the effective mitigation of the integrated perturbation amplitude. In this paper we employ this phase control mechanism to mitigate the sausage and kink instability growths.

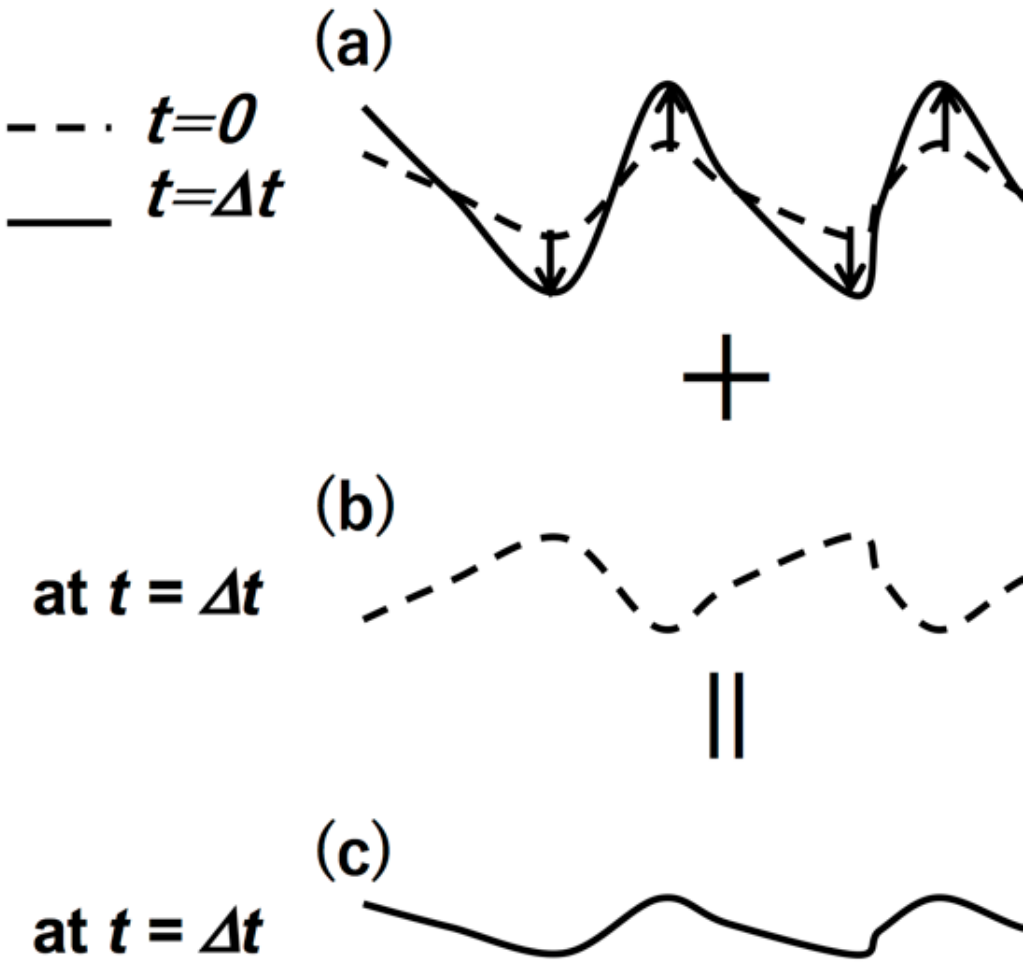

Fig. 1 Dynamic mitigation mechanism. (a)At $t$=0 a perturbation is imposed. The initial perturbation grows with $\gamma$ for an unstable system. (b) After $\Delta t$ another perturbation, which has an inverse phase, is actively imposed. (c) After the superimposition of the perturbations in (a) and (b) at $\Delta t$, the actual perturbation amplitude is mitigated well.

First, we apply the phase control mechanism shown in Fig. 1 to the sausage instability of a magnetized $z$-current driven plasma column. The column plasma consists of protons and electrons. The protons are stationary and uniform in space inside the column. Its initial density is $f \times n_0$, and the protons provide the partial charge neutralization of the $z$-current electrons with the neutralization ratio of *f*. In this paper *f* is set to be 0.99. The electron beam speed in *z* is $0.1c$, and the electron number density is $n_0$. Here $c$ is the speed of light and $n_0 = 10^{16}$/m$^3$. The electron velocity and density are also uniform. The initial ion and electron temperatures are 1eV. The plasma column is 20 cm long in *z*, and its radius is 1.5 cm. In the *z* direction, the cyclic boundary condition is employed, and in the transverse directions the open boundary conditions are employed. In the column equilibrium state the radial force is balanced between the outward electrostatic force and the inward magnetic pinching force by the azimuthal magnetic field generated by the electron net current[6]. The 3D particle-in-cell simulations are performed by EPOCH3D[29].

Figures 2 present the sausage instability evolution without the mitigation mechanism. Figures 2 demonstrate a clear instability growth, and Figs. 2 show the electron distributions at (a)*t*=20 ns and (b)*t*=30ns and the proton distributions at (c)*t*=20ns and (d)*t*=40ns. Figures 3 shows the electron densities at (a)*t*=0 ns and (b)*t*=36ns and the proton distribution at (c)*t*=40ns. As presented in Figs. 2 the initial perturbation has 4 modes in the 20cm simulation box in the longitudinal direction of *z*. In Figs. 3, an electron beam rotation is added to smooth the initial perturbation for the sausage instability by one rotation in 20cm with 20% amplitude of the initial plasm column amplitude. Figure 3(a) shows the rotation profile, and the stationary ion column has no perturbation initially. Compared with Fig. 2(d), Fig. 3(c) demonstrates that the sausage instability is well mitigated by the rotation behavior imposed. Figure 4 shows the histories of the normalized field energy, which is proportional to $\varepsilon_0 \vec{E}^2 + \mu_0 \vec{B}^2$, for the sausage instability presented in Figs. 2 and 3. As explained above, the rotating electron driver beam smooth and mitigate the sausage instability growth clearly. The growth mitigation is significant and the growth mitigation ratio is 58.3% at *t*=40nsec. In the growth mitigation mechanism as mentioned earlier, the growth rate does not change but the perturbation amplitude is reduced significantly. Figures 2-4 demonstrate that the growth amplitude reduction is remarkable for the sausage instability.

Figures 5 show the kink instability growth without the mitigation mechanism, and initially the plasma electron column is displaced in *y* by 5% of the column radius. In 20 cm 2 waves is accommodated in *z*. The electron distributions are presented in Figs. 5(a) at *t*=0ns and (b) *t*=50ns, and the protons are shown in Figs. 5(c) at *t*=0ns and (d) *t*=50ns. Figures 5 (b) and (d) present the kink instability growth clearly. Figures 6 present the kink instability growth under the phase control mitigation of 4 rotations in 20 cm in $z$ with the amplitude of 20% of the column radius. The electrons are presented in Figs. 6(a) at *t*=8ns and (b) *t*=50ns, and the protons are shown in Figs. 6(c) at *t*=14ns and (d) *t*=50ns. The electron and proton spatial distributions reflect the electron beam rotation around the column axis in Figs. 6. The kink instability shown in Figs. 5 is well mitigated by the rotation behavior as presented in Figs. 6. Figure 7 shows the histories of the normalized field energy for the kink mode. The spiral motion of the electron beam additionally induces a spiral kink instability (see Figs. 6(a) and (c)) especially at the early stage of the perturbation growth. The growth rate $\gamma$ of the kink instability is proportional to $\gamma \propto (B_\theta k)/(\mu_0 \rho)$, where $B_\theta$ is the azimuthal magnetic field, $k$ the wave number and $\rho$ the mass density [7, 27]. Figures 6(b) and (d) demonstrate that the mitigation mechanism driven by the driver electron beam rotation smooths the initial perturbations. Figure 7 also presents the fact, and initially the spiral kink mode emerges first, because of its higher mode. However, after that, Fig. 7 demonstrates that the spiral perturbation mitigates the initial displacement in $y$ well by the spiral motion. Finally, the overall perturbation growth is mitigated. The mitigation ratio is 30.0% at *t*=50nsec.

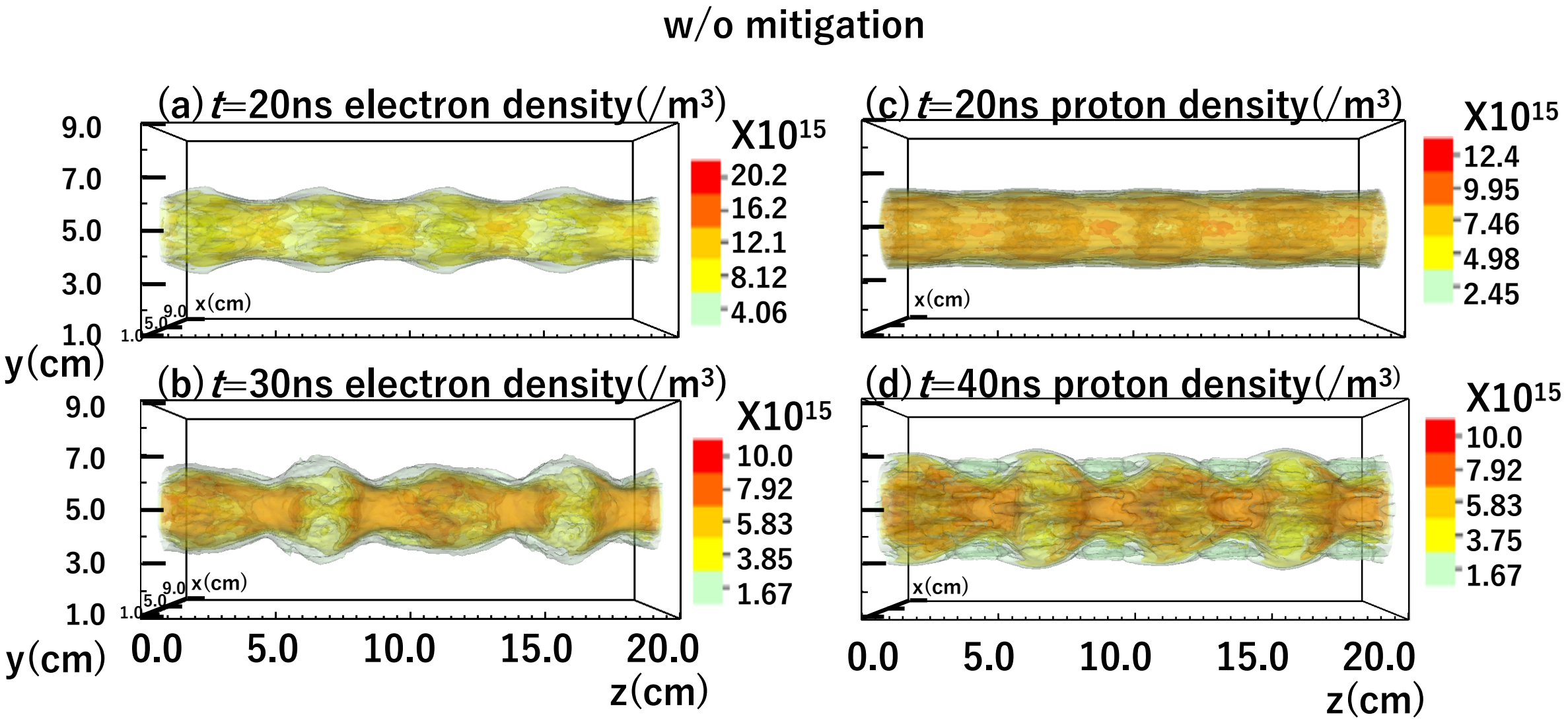

Fig. 2 Sausage instability without the mitigation mechanism. The electron densities at (a) *t*=20ns and (b) *t*=30ns and the proton densities at (c) *t*=20ns and (d) *t*=40ns.

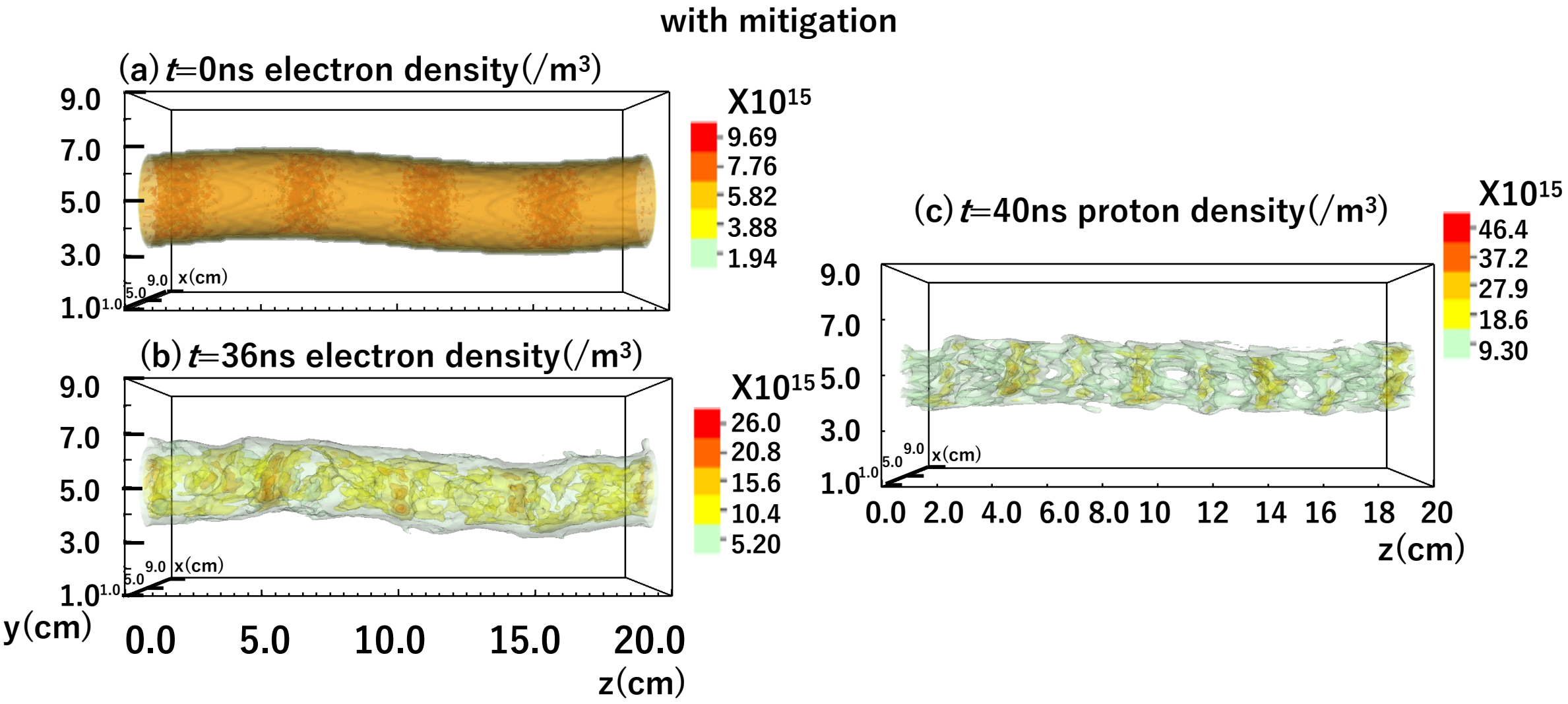

Fig. 3 Sausage instability with the mitigation mechanism. The electron densities at (a) $t$=0ns and (b) $t$=36ns and the proton density at (c) $t$=40ns.

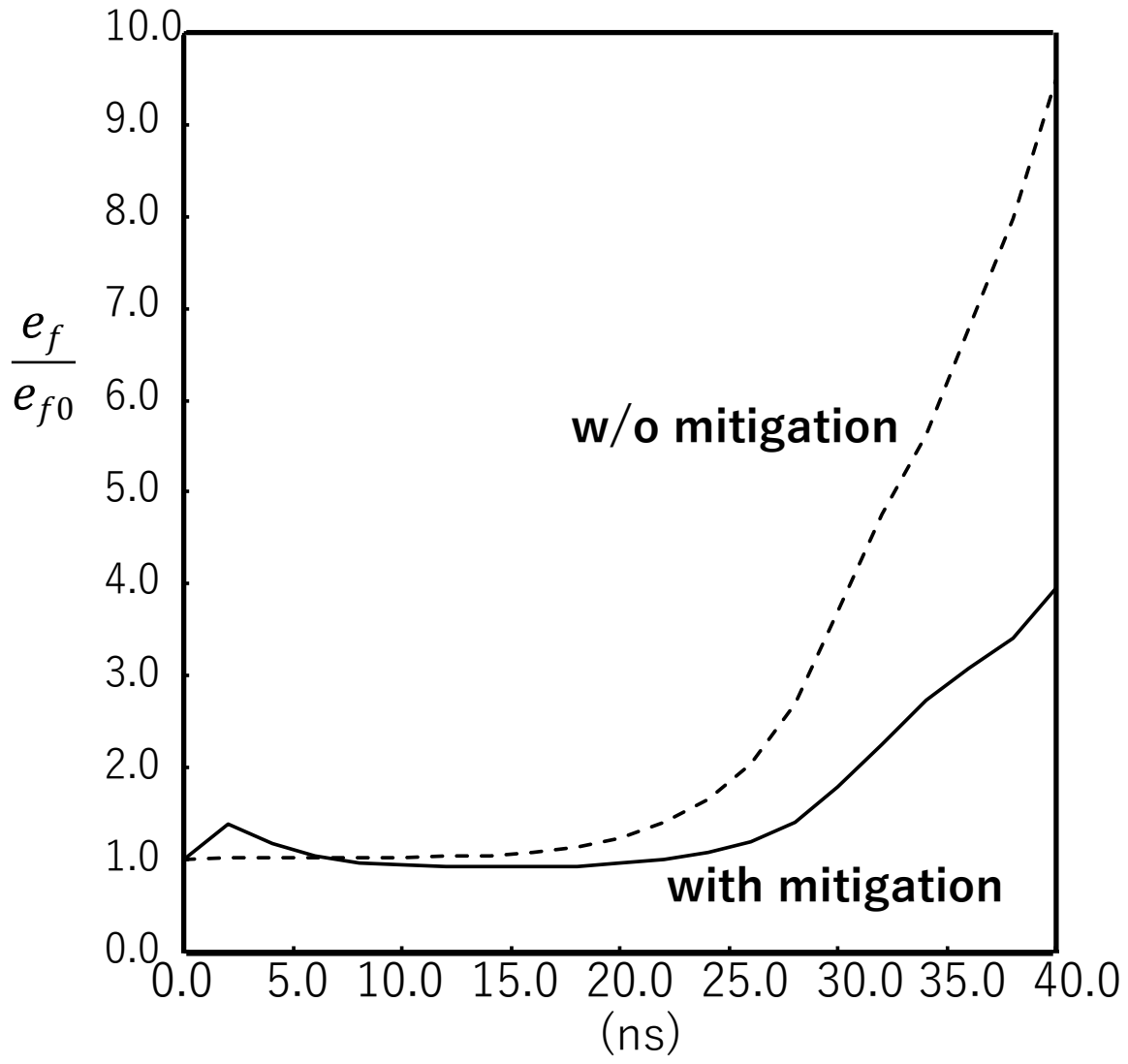

Fig. 4 Field energy histories for sausage instability with the mitigation phase control (solid line) and without the control (dotted line).

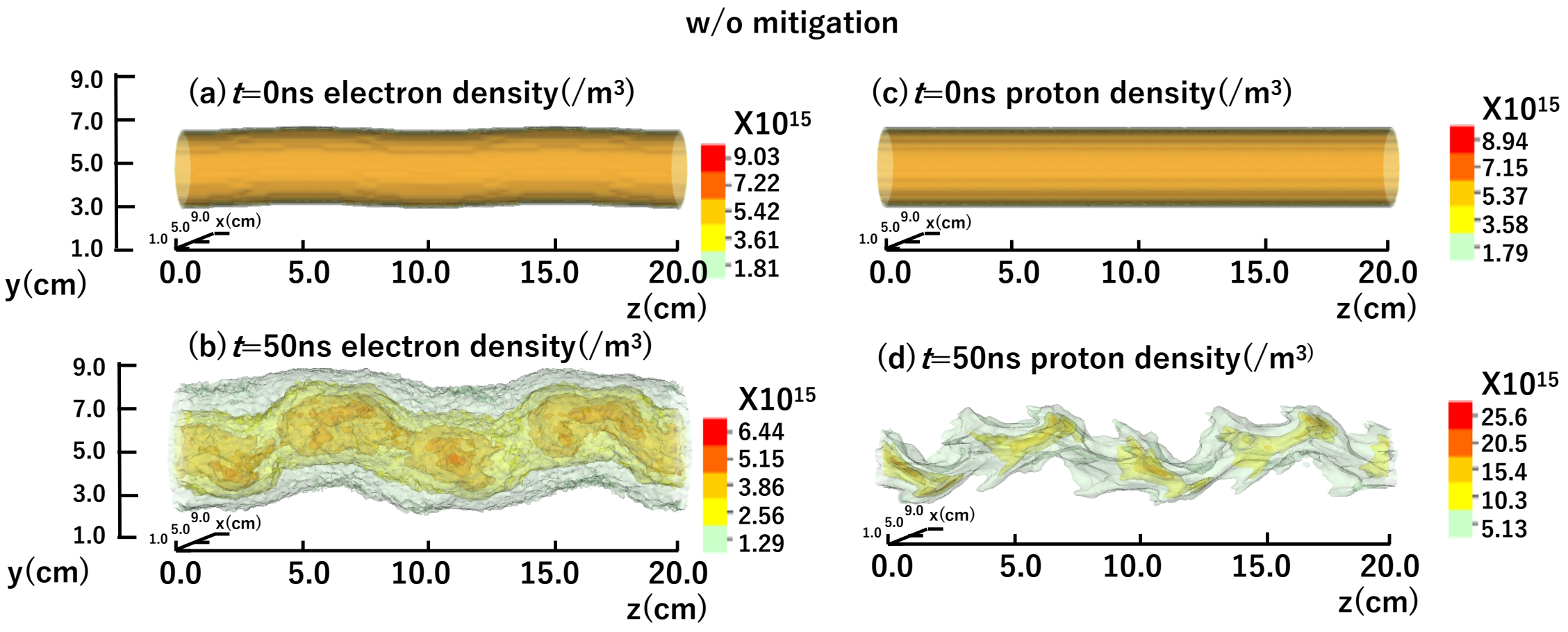

Fig. 5 Kink instability without the mitigation mechanism. The electron densities at (a) $t$=0ns and (b) $t$=50ns and the proton densities at (c) $t$=0ns and (d) $t$=50ns.

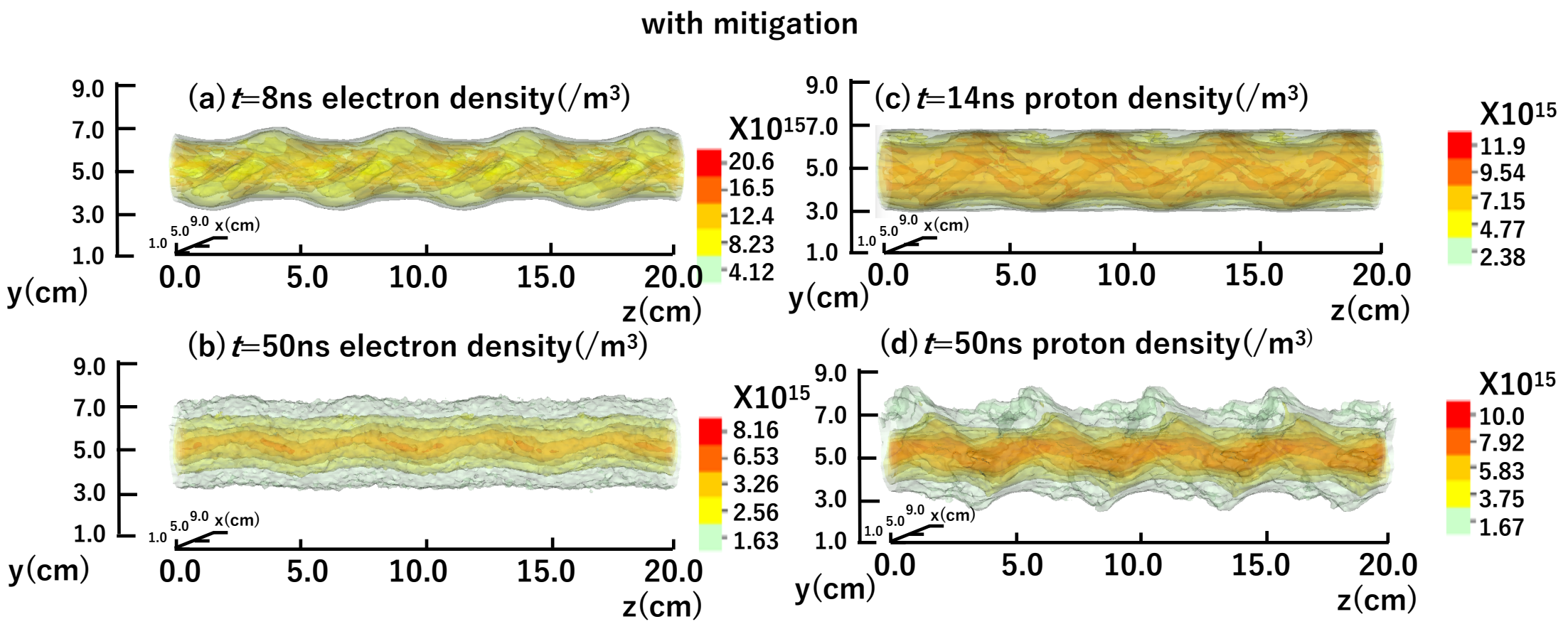

Fig. 6 Kink instability with the mitigation mechanism. The electron densities at (a) $t$=8ns and (b) $t$=50ns and the proton density at (c) $t$=14ns and (d) $t$=50ns.

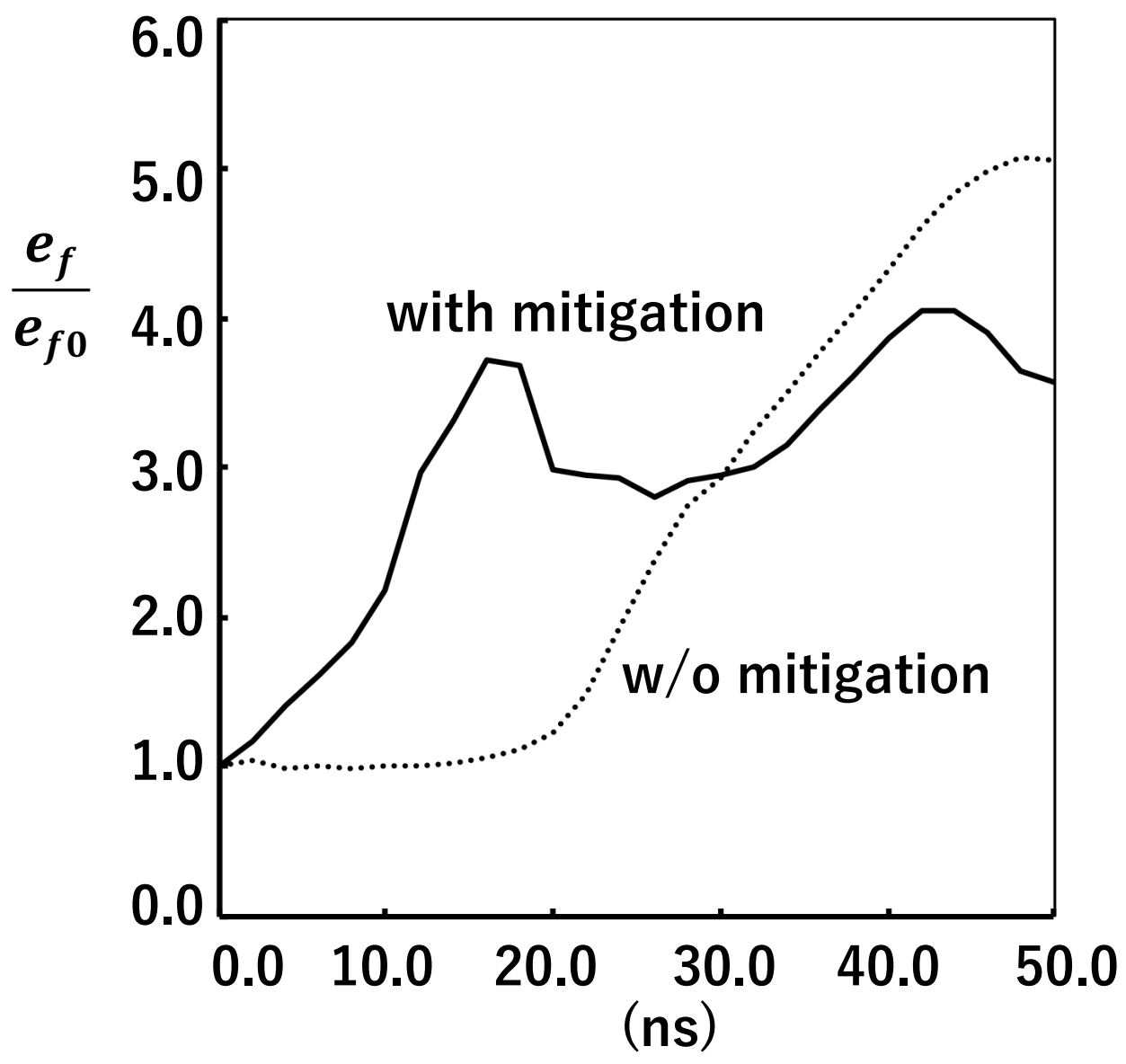

Fig. 7 Field energy histories for kink instability with the mitigation phase control (solid line) and without the control (dotted line).

We also have to point out that the dynamic mitigation mechanism is not almighty. If the non-uniformity phase cannot be controlled actively, the mechanism cannot be realized. For example, if the initial plasma column has the initial perturbation, which cannot be controlled from the outside, the dynamic mitigation mechanism does not work. In addition, it would be better to point out again that the growth rate $\gamma$ of the plasma instability does not change as discussed above. The results in this paper come from the mitigation of the perturbation amplitude.

The paper presented the dynamic mitigation of the instabilities for the plasma column. The new dynamic mitigation method for plasmas comes from the control theory[11]. The perfect feedback control cannot be realized in plasmas. However, if we can actively oscillate the perturbation phase, the dynamic mitigation mechanism is realized[1-5]. The dynamic mitigation mechanism would be a kind of the feed forward control[11]. In the paper the dynamic mitigation mechanisms works well to mitigate the sausage and kink instabilities of the z-current driven plasma column.

## ACKNOWLEDGEMENTS

The work was partly supported by Japan Society for the Promotion of Science (JSPS), Ministry of Education, Culture, Sports, Science and Technology (MEXT), Japan / U. S. Cooperation in Fusion Research and Development, Center for Optical Research and Education, Utsunomiya University (CORE), and Institute of Laser Engineering, Osaka University (ILE).